\documentclass[aps,twocolumn]{revtex4-1}
\usepackage{amsmath,amssymb,graphicx,enumerate}

\newcommand{\amp}{&}
\newcommand{\mpl}{M_{Pl}}
\newcommand{\beq}{\begin{equation}}
\newcommand{\eeq}{\end{equation}}
\newcommand{\bea}{\begin{eqnarray}}
\newcommand{\eea}{\end{eqnarray}}
\newcommand{\bp}{\lambda}
\newcommand{\dr}{\Gamma_\phi}
\newcommand{\drreq}{\Gamma_{\phi,\mbox{\tiny{req}}}}
\newcommand{\as}{A}
\newcommand{\Nphi}{N_\phi}
\newcommand{\Nphib}{N_{\bar{\phi}}}
\newcommand{\nphi}{n_\phi}
\newcommand{\nphib}{n_{\bar{\phi}}}
\newcommand{\Nb}{N_b}
\newcommand{\Nbb}{N_{\bar{b}}}
\newcommand{\nb}{n_b}
\newcommand{\nbb}{n_{\bar{b}}}
\newcommand{\bphi}{b_\phi}
\newcommand{\gamf}{\gamma}
\newcommand{\fud}{\beta}

\begin{document}

\title{Generating the Observed Baryon Asymmetry from the Inflaton Field}

\begin{abstract}
We propose a  mechanism by which the inflaton can generate baryogenesis, by taking the inflaton to be a complex scalar field with a weakly broken global symmetry and present a new version of the Affleck-Dine mechanism. The smallness of the breaking is motivated both by technical naturalness and a requirement for inflation. We study inflation driven by a quadratic potential for simplicity and discuss generalizations to other potentials. We compute the inflationary dynamics and find that a conserved particle number is obtained towards the end of inflation. We then explain in detail the later decay to baryons. We present two promising embeddings in particle physics: (i) using high dimension operators for a gauge singlet; we find this leads to the observed asymmetry for decay controlled by the $\sim$ grand unified theory scale and this is precisely the regime where the effective field theory applies. (ii) using a colored inflaton, which requires small couplings. We also point out two observational consequences: a possible large scale dipole in the baryon density, and a striking prediction of isocurvature fluctuations whose amplitude is found to be just below current limits and potentially detectable in future data.
\end{abstract}

\author{Mark P.~Hertzberg$^*$ and Johanna Karouby$^\dagger$}
\affiliation{Center for Theoretical Physics and Dept.~of Physics\\
Massachusetts Institute of Technology, Cambridge, MA 02139, USA}

\date{\today}

\vspace*{-14.2cm} {\hfill MIT-CTP 4493\,\,\,\,} 

\maketitle

\tableofcontents

\section{Introduction}

The Standard Model of particle physics is a major success of modern physics, accurately describing all laboratory observations, and recently confirmed by LHC data \cite{ATLAS,CMS}.
On the other hand, there are several cosmological observations that are left unaccounted for.
One is the asymmetry between matter and anti-matter. 

Cosmological observations in the CMB, and other measurements, have revealed that the asymmetry between matter and anti-matter is small. It is normally quantified by the parameter $\eta$ which is the average baryon-to-photon ratio in the universe, with the small value
\beq
\eta_{obs}\approx 6\times 10^{-10}.
\eeq
As discovered by Sakharov \cite{Sakharov:1967dj}, any particle physics mechanism that seeks to account for this, starting from a state of symmetry, requires (i) the violation of baryon number, (ii) the breaking of C and CP symmetry, and (iii) out of equilibrium processes. These conditions are not effectively satisfied by the Standard Model of particle physics, strongly suggesting new physics. 

One might try to avoid this problem by simply imagining that the universe began with the asymmetry. However, such a proposal appears both unsatisfying and unlikely due to cosmological inflation; a phase of exponential expansion in the early universe \cite{Guth,Linde}. Evidence is continually mounting, including recent CMB data \cite{Hinshaw:2012aka,Ade:2013uln}, that this paradigm explains not only the large scale homogeneity and isotropy, but also the density inhomogeneities and and temperature anisotropies.
 An important consequence is that such an exponentially growing phase would wipe out any initial baryon number. 

Ordinarily, to avoid this problem due to cosmological inflation, one introduces a collection of new fields that enter well after the inflationary phase in order to generate a baryon number. There exists many proposals for baryogenesis of this sort; for reviews see, e.g., \cite{Riotto:1998bt,Cline:2006ts,Buchmuller:2007fd}. One popular mechanism is to introduce new physics around the electroweak scale and achieve baryogenesis at the electroweak phase transition, e.g., see \cite{Trodden:1998ym,Menon:2009mz,Cline:2002aa,Rius:1999zc}. As we have yet to see any new physics at the LHC, which is probing this energy regime, it is important to consider the possibility that baryogenesis is associated with much higher energies. Perhaps the only known probe to physics at very high energies is through inflation.

In this paper, and accompanying letter \cite{BaryogenesisLetter}, we show that although inflation wipes out any initial matter/anti-matter asymmetry, the asymmetry can still be generated by the inflaton itself. The key reason this is possible is that the inflaton acquires a type of vev during inflation and this information is not wiped out by the inflationary phase. In order to connect this to baryogenesis, we will put forward a new variation on the classic Affleck-Dine \cite{AffleckDine} mechanism for baryogenesis, which uses scalar field dynamics to obtain a net baryon number. In the original proposal, Affleck-Dine used a complex scalar field, usually thought to be unrelated to the inflaton but possibly a spectator field during inflation, to generate baryons in the radiation or matter eras. 

Many interesting versions and constraints, often including connections to supersymmetry, have been found for these types of Affleck-Dine models; some interesting works appears in Refs.~\cite{Allahverdi:2012ju,Dine:1995kz,Dine:1995uk,Charng:2008ke,Enqvist:1998pf,Mazumdar:2001nw,Koyama:1998hk,Allahverdi:2001is,Seto:2005pj,Kasuya:2006wf,Dutta:2010sg,Marsh:2011ud,Garcia:2013bha,Enqvist1998,Delepine2007,BasteroGil:2011cx,Kitano:2008tk,Murayama:1993xu,Murayama:1992ua}. For instance, Ref.~\cite{Kitano:2008tk} considers an interesting leptogenesis model. 
In Ref.~\cite{Mazumdar:2001nw} an interesting extra-dimension model was put forward. In Ref.~\cite{Allahverdi:2001is} radiative corrections were considered. In Ref.~\cite{Seto:2005pj} connections to graviton dark matter were investigated. In Ref.~\cite{Garcia:2013bha} corrections from moduli were computed.
Many works have been in the context of supergravity, such as the very interesting work in Ref.~\cite{Charng:2008ke}, which has overlap with our work here. However, with no current evidence for (low scale) supersymmetry, it is very useful to consider less restrictive frameworks. Here we provide a general model using only the tools of effective field theory, constrained by the latest cosmological data. Our model will be minimal and predictive.

Here we propose a new model where the aforementioned complex scalar field is the inflaton itself. We study both particle physics and cosmological aspects of our model, including current observational constraints. This work is a full treatment of the basic idea summarized in our accompanying letter \cite{BaryogenesisLetter}.
Our key ideas and findings are summarized as follows:
\begin{itemize}
\item We propose that the inflaton is a complex scalar field with a weakly broken global $U(1)$ symmetry. For simplicity, we consider inflation driven by a symmetric quadratic potential, plus a sub-dominant symmetry breaking term.
\item We find that a non-zero particle number is generated in the latter stage of inflation. After inflation this can decay into baryons and eventually produce a thermal universe.
\item We propose two promising particle physics models for both the symmetry breaking and the decay into baryons:
\begin{enumerate}[(i)]
\item Utilizing high dimension operators, which is preferable if the inflaton is a gauge singlet.
\item Utilizing low dimension operators, which is natural if the inflaton carries color. 
\end{enumerate}
\item We find that model (i) predicts the observed baryon asymmetry if the decay occurs through operators controlled by $\sim$\,GUT scale and this is precisely the regime where the EFT applies.
\item We find that model (ii) requires small couplings to obtain the observed baryon asymmetry; which may be interesting in supersymmetric contexts.
 \item We find that the model allows a large scale baryon dipole in the universe, whose amplitude depends on the number of e-foldings of inflation.
 \item We find that the model predicts a baryon isocurvature fluctuation at a level consistent with the latest CMB bounds, and potentially detectable in future generations of data.
\end{itemize}

The basic outline of this paper is as follows: 
In Section \ref{Model} we specify the model and computational techniques, 
in Section \ref{Results} we obtain results for the scalar field asymmetry, 
in Section \ref{Decay} we discuss the decay into baryons, 
in Section \ref{Inflation} we discuss constraints from inflation, 
in Section \ref{Particle} we discuss particle physics realizations, 
in Section \ref{Observational} we mention possible observational consequences,
in Section \ref{Conclusions} we discuss our results,
and in the Appendix we present further analytical results.

\section{Complex Scalar Model}\label{Model}

Consider a complex scalar field $\phi$, coupled to gravity, with dynamics governed by the standard two-derivative action 
(signature $+ - - -$, units $\hbar=c=1$)
\beq
S=\int\! d^4x\sqrt{-g}\left[{1\over 16\pi G}\mathcal{R}+|\partial\phi|^2-V(\phi,\phi^*)\right].
\eeq
With a canonical kinetic sector, our entire freedom comes from the specification of the potential function $V$. For our purposes, it will be useful to decompose the potential into a ``symmetric" piece $V_s$ and a ``breaking" piece $V_b$ piece, with respect to a global $U(1)$ symmetry $\phi\to e^{-i\alpha}\phi$
\beq
V(\phi,\phi^*)=V_s(|\phi|)+V_b(\phi,\phi^*).
\eeq
In order to describe inflation we assume that the symmetric piece $V_s$ dominates, even at rather large field values where inflation occurs.
By expanding $\phi$ around 0, we assume that the symmetric piece contains a quadratic mass term 
\beq
V_s(|\phi|)=m^2|\phi|^2,
\eeq
which clearly respects a $\phi\to e^{-i\alpha}\phi$ global symmetry. 
It is well known that a purely quadratic potential (without higher order corrections to $V_s$) will establish large field, or ``chaotic" inflation \cite{Linde83}.
This is a simple model of inflation that will provide a useful pedagogical tool to describe our  mechanism for baryogenesis. As we will discuss in Section \ref{Inflation}, such a model is in good agreement with the spectrum of density fluctuations in the universe, though it is in small tension with recent Planck constraints on tensor fluctuations \cite{Ade:2013uln}. However, generalizing to other symmetric inflationary potentials is straightforward, such as models that ``flatten" at large field values.

The global symmetry is associated with a conserved particle number. So to generate a non-zero particle number (that will decay into baryons) we add a higher dimension operator that explicitly breaks the global $U(1)$ symmetry
\beq
V_b(\phi,\phi^*)=\bp(\phi^n+\phi^{*n}),
\label{br}
\eeq
with $n\ge3$. Note that other terms, such as $\sim(\phi^{n-1}\phi^*+\phi^{*n-1}\phi)$ would also break this $U(1)$ symmetry. Since we do not expect any qualitative differences to our central results, we restrict ourselves to the simple symmetry breaking potential given in eq.~(\ref{br}).
Another motivation of focussing purely on this simple potential is that it is protected by a discrete $\mathbb{Z}_n$ symmetry of the form $\phi\to e^{2\pi i/n}\phi$. This symmetry forbids the generation of many other operators, such as $\sim(\phi^2+\phi^{*2})$, which would cause the $U(1)$ symmetry to be broken even at low energies. Instead the $\mathbb{Z}_n$ protects the structure of the Lagrangian under RG flow. Possible corrections that can be generated include $\sim\lambda^2(\phi^{2n}+\phi^{*2n})/\mpl^4$, which is entirely negligible in our regime of interest.
In short, our model includes the set of leading operators that respect a discrete $\mathbb{Z}_n$ symmetry and weakly breaks a global continuous $U(1)$ symmetry.

We assume that the breaking parameter $\bp$ is very small (in appropriate units to be discussed in Section \ref{Inflation}) so that the global symmetry is only weakly broken.
This assumption of very small $\bp$ is motivated by two main reasons:
\begin{enumerate}
\item Since $\bp$ is responsible for the breaking of a symmetry, it is technically natural for it to be small according to the principles of effective field theory.
\item The smallness of $\bp$ is an essential requirement on any inflationary model so that such higher order corrections do not spoil the flatness of the potential $V_s$. (The breaking term  adds steep positive and negative parts to the potential in the complex plane, unless $\bp$ is small.)
\end{enumerate}
As should be expected, the smallness of $\bp$ favors a small baryon-to-photon ratio, as we examine in detail later.

Although the global symmetry $\phi\to e^{-i\alpha}\phi$ is explicitly broken by the higher dimension operator $\bp(\phi^n+\phi^{*n})$,
this theory does respect the $\phi\leftrightarrow\phi^*$ symmetry, which is the charge conjugation symmetry. However, in order to satisfy the Sakharov conditions for baryogenesis, this charge conjugation (and CP) symmetry can be broken spontaneously. Affleck-Dine \cite{AffleckDine} assumed it was broken spontaneously by some light field that acquires a vev in the radiation or matter eras. 

In this work, we identify $\phi$ with the inflaton and use the inflationary phase to obtain a vev for $\phi$. As a result we are unifying inflation with baryogenesis.

\subsection{Particle/Anti-Particle Asymmetry}\label{Asymmetry}

For the above class of models, we would like to compute the details of the inflationary phase, the post-inflationary phase, and ultimately the transfer of energy to radiation including baryons. To begin, we note that since $n\geq3$, then at late times the inflaton $\phi$ becomes small, the $\phi\to e^{-i\alpha}\phi$ symmetry violating term becomes negligible, and the symmetry becomes respected. By Noether's theorem this is associated with a conserved particle number
\beq
\Delta \Nphi=\Nphi-\Nphib=i\int d^3x\sqrt{g_s}\,(\phi^*\dot\phi-\dot\phi^*\phi),
\eeq
where $d^3x\sqrt{g_s}$ is the spatial volume measure. $\Nphi$ ($\Nphib$) is the number of $\phi$-particles (anti-particles), and
later we will relate $\Delta\Nphi$ to baryon number.
Since $\phi$ is taken to be the inflaton, then we know to excellent approximation that $\phi$ is homogeneous on large scales. So in an FRW universe, the spatial integral can be immediately performed in terms of some comoving volume $V_{com}$ and scale factor $a(t)$, giving
\beq
\Delta \Nphi=\Nphi-\Nphib=i\, V_{com}\,a^3(\phi^*\dot\phi-\dot\phi^*\phi).
\eeq

\begin{figure}
\center{\includegraphics[width=4.7cm]{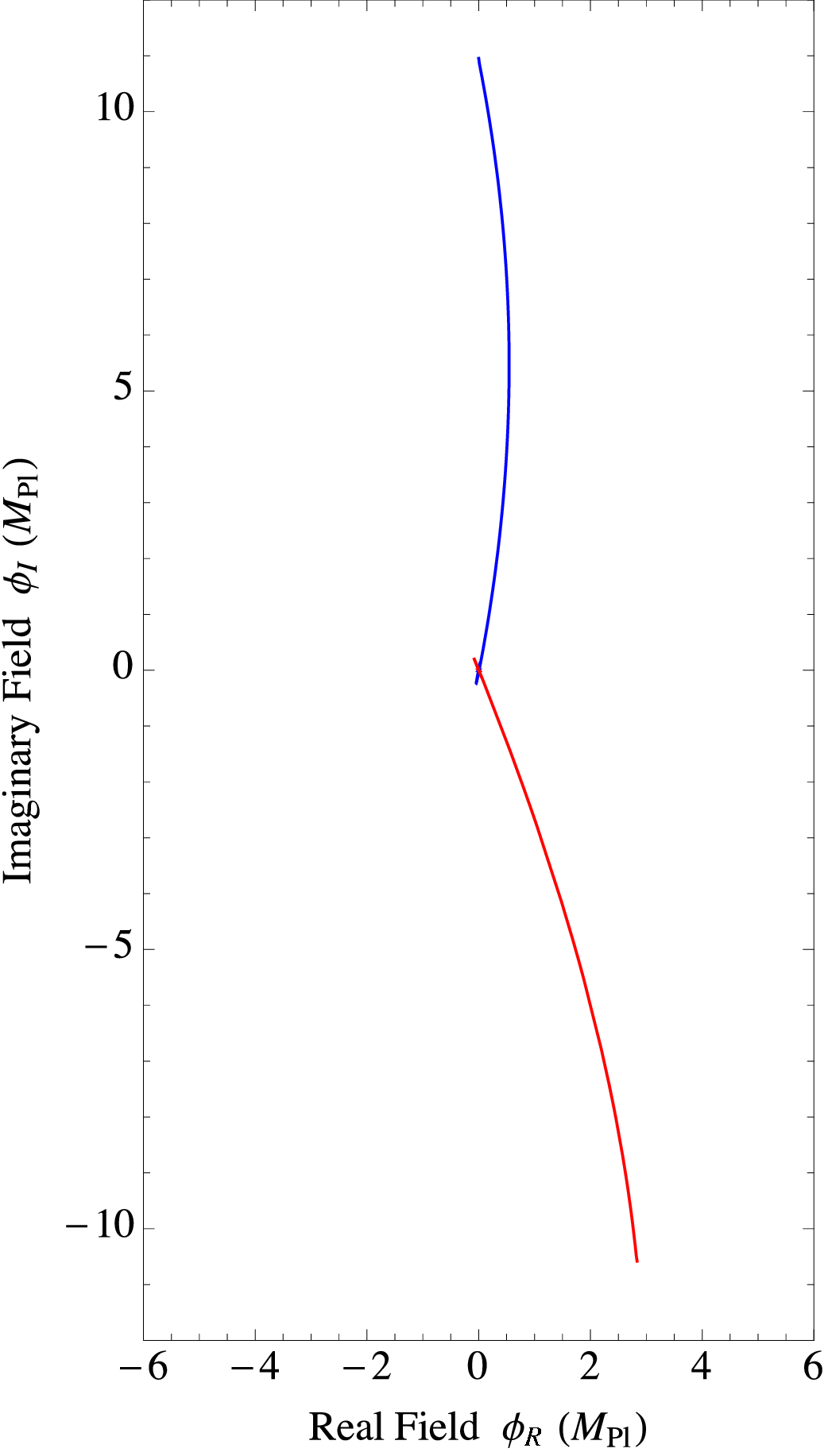}
\includegraphics[width=3.65cm]{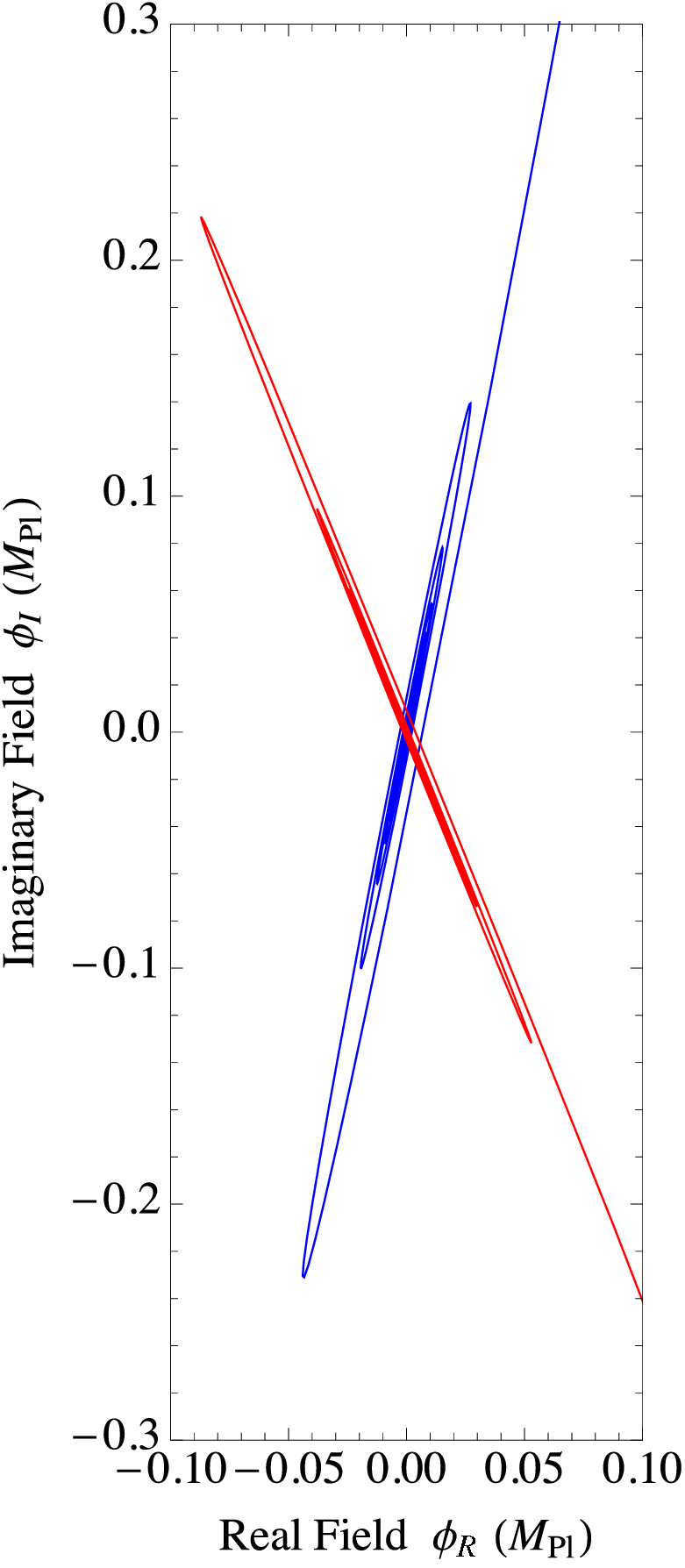}}
\caption{Field evolution in the complex $\phi$-plane for $n=3$ and $\bp\mpl/m^2=0.006$, with initial condition $\rho_i=2\sqrt{60}\,\mpl$. Left is zoomed out and shows early time behavior during slow-roll inflation. Right is zoomed in to $\phi=0$ and shows late time elliptic motion. Blue (upper) curve is for initial angle $\theta_i=\pi/2$ and red (lower) curve is for initial angle $\theta_i=-5\pi/12$.}
\label{FieldEvolution}\end{figure}

To be self consistent we ignore spatial gradients, and the equation of motion for $\phi$ is
\beq
\ddot\phi+3 H\dot\phi+m^2\phi+\bp\,n\,\phi^{* n-1}=0,
\eeq
where $H=\dot{a}/a$ is the Hubble parameter.
The evolution of the field, including the early time behavior during slow-roll inflation and the late time behavior as the field acquires elliptic motion around the origin, is shown in Fig.~\ref{FieldEvolution} for two different initial conditions.

By taking a time derivative of $\Delta \Nphi$ and using the equation of motion, it is simple to obtain an alternate expression for $\Delta \Nphi$ at some final time $t_f$ in terms of its value at some initial time $t_i$
\bea
\Delta \Nphi(t_f) \amp=\amp \Delta \Nphi(t_i)\nonumber\\
\amp+\amp i\, \bp\, V_{com}\,n\int _{t_i}^{t_f} dt\,a(t)^3(\phi(t)^n-\phi^*(t)^{n}).\,\,\,\,\,
\label{dN1}\eea
It is appropriate that the integral has a prefactor of $\bp$, as the particle number must be conserved in the $\bp\to 0$ limit.

We now rewrite the complex field in polar co-ordinates
\beq
\phi(t)={1\over\sqrt{2}}\rho(t)\,e^{i\theta(t)}.
\eeq
Substituting into eq.~(\ref{dN1}) leads to
\bea
\Delta \Nphi(t_f) \amp=\amp \Delta \Nphi(t_i)\nonumber\\
\amp-\amp \bp{V_{com}\,n\over 2^{{n\over2}-1}}\int _{t_i}^{t_f} dt\,a(t)^3\rho(t)^n\sin(n\,\theta(t)).\,\,\,\,\,\,\,
\label{dNint}\eea
We will find this integral representation to be very convenient, as we explain in the next subsection.

\subsection{Small Coupling Approximation}\label{SmallGamma}

  At leading order in $\bp$ we can reduce the complexity of the problem significantly. Since the expression in eq.~(\ref{dNint}) already has a factor of $\bp$ in front of the integral, then to first order in $\bp$ we only need to evaluate the quantities in the integral to zeroth order in $\bp$.

By assuming the field begins deep in the inflationary era, then any initial number of $\phi$ particles are diluted by inflation and so we can ignore $\Delta \Nphi(t_i)$.  Then as the field evolves at zeroth order in $\bp$, it evolves {\em radially} in the complex $\phi$ plane. This means that we can simply take $\theta(t)=\theta_i$ as a constant, and allow $\rho(t)$ to oscillate, i.e., we allow $\rho(t)$ to take the sign of the field, either positive or negative. This reduces the problem significantly to solving only a single ordinary differential equation. At first order in $\bp$, $\Delta \Nphi$ is simply
\beq
\Delta \Nphi(t_f) = - \bp{V_{com}\,n\over 2^{{n\over2}-1}}\sin(n\,\theta_i)\,I(t_i,t_f),
\label{deltaNsimp}\eeq
where
\beq
I(t_i,t_f)\equiv\int _{t_i}^{t_f} dt\,a(t)^3\rho_0(t)^n.
\eeq
Here $\rho_0$ is a real valued function satisfying the quadratic potential version of the equation of motion
\beq
\ddot\rho_0+3 H_0\dot\rho_0+m^2\rho_0=0,
\label{rhosimp}\eeq
with corresponding Friedmann equation (we assume flat FRW)
\beq
H_0^2={\varepsilon_0\over 3\mpl^2},\,\,\,\,\,\,\,\,\,\,\,\varepsilon_0={1\over2}\dot\rho_0^2+{1\over2}m^2\rho_0^2,
\label{Hsimp}\eeq
where $\mpl\equiv1/\sqrt{8\pi G}$ is the reduced Planck mass. So by solving for a single degree of freedom in a quadratic potential, we have an expression for the particle number in the small $\bp$ regime.

It is worthwhile to note that for particular values of the initial angle $\theta_i$, such that $\theta_i={p\pi\over n}\,|\,p\in\mathbb{Z}$, no asymmetry is generated due to the $\sim\sin(n\,\theta_i)$ factor. In this case, the classical motion of the field is exactly radial as we can assume $\dot{\theta}_i=0$ (non-zero $\dot\theta$ is quickly erased by Hubble friction). Near these special values of the initial angle there is a large isocurvature fluctuation (see eq.~(\ref{deltaeta})). In the following, since we are interested in baryogenesis, we consider $\theta_i$ to be a typical generic value rather than these special ones.

\subsection{Dimensionless Quantities}

Although $\Delta \Nphi$ is dimensionless, it is extrinsic, depending on the size of the universe. It is useful to define a related intrinsic quantity, namely the particle density
\beq
\Delta \nphi={\Delta \Nphi\over V_{com}\,a^3}.
\label{density}\eeq
But this is now dimensionful. In order to obtain a dimensionless, intrinsic measure of asymmetry, we divide by the energy density $\varepsilon_0$
\beq
\as\equiv{m\,\Delta \nphi\over \varepsilon_0}.
\label{ratio}\eeq
This is appropriate because at late times the energy density is provided by a gas of non-relativistic $\phi$ and anti-$\phi$ particles with energy density
\beq
\varepsilon_0 = m (\nphi+\nphib).
\eeq
This means that at late times, the dimensionless asymmetry variable $\as$ reduces to
\beq
\as={\Delta \nphi\over \nphi+\nphib}={\nphi-\nphib\over \nphi+\nphib},
\label{asdef}\eeq
which is clearly a good measure of the asymmetry.

Moreover, we would like to introduce dimensionless variables. We introduce a dimensionless time variable $\tau$, a dimensionless field variable $\bar{\rho}$, and a dimensionless Hubble parameter $\bar{H}$ as follows
\beq
\tau \equiv m\, t,\,\,\,\,\,\,\,\,\,\,\bar{\rho}\equiv{\rho_0\over\mpl},\,\,\,\,\,\,\,\,\,\,\bar{H}\equiv {H_0\over m}.
\eeq
In terms of these new variables we can express $\as$ as
\beq
\as = -f_n(\tau_i,\tau_f){\bp\mpl^{n-2}\over m^2}\sin(n\,\theta_i),
\label{Aexp}\eeq
where the dimensionless function $f_n$ given by
\beq
f_n(\tau_i,\tau_f)={n\over 2^{{n\over 2}-1}}{\bar{I}(\tau_i,\tau_f)\over a(\tau_f)^3\,\bar{\varepsilon}(\tau_f)},
\label{fdef}\eeq
where
\beq
\bar{I}(\tau_i,\tau_f)\equiv\int_{\tau_i}^{\tau_f}d\tau\,a(\tau)^3\bar{\rho}(\tau)^n.
\label{barI1}\eeq
The dimensionless field $\bar{\rho}$ satisfies the dimensionless version of the differential equation (\ref{rhosimp}).
While $a(\tau),\,\bar{\varepsilon}$ satisfy the dimensionless version of the Friedmann equation (\ref{Hsimp}).
We have now extracted all the dimensionful parameters of the theory, leaving the task of solving for $f_n$.
In order to do so, we perform numerics and then give an analytical estimates. From now on, we only use the dimensionless variables introduced in this section.

\section{Asymmetry Results}\label{Results}

\subsection{Numerical Findings}\label{Numerical}

Although we cannot exactly solve this nonlinear differential equation analytically in general, we can solve it numerically. (See the Appendix for exact analytical results in the Affleck-Dine regime).

In the limit in which we take $\tau_i$ very early during slow-roll inflation and we take $\tau_f$ very late after inflation (a type of $\phi$-matter dominated era), then $f_n$ becomes independent of both $\tau_i$ and $\tau_f$. It becomes a constant only dependent on the power $n$. 
The reason is the follows: At very early times during slow-roll inflation, $a(\tau)$ is exponentially small, so the lower part of the integral is negligible. At very late times during the $\phi$-matter dominated era, $\bar{\rho}$ redshifts away, so the upper part of the integral becomes negligible. Instead only the ``middle" part of the integral is important; the latter stage of inflation just before the transition to matter. 
\begin{figure}
\center{\includegraphics[width=\columnwidth]{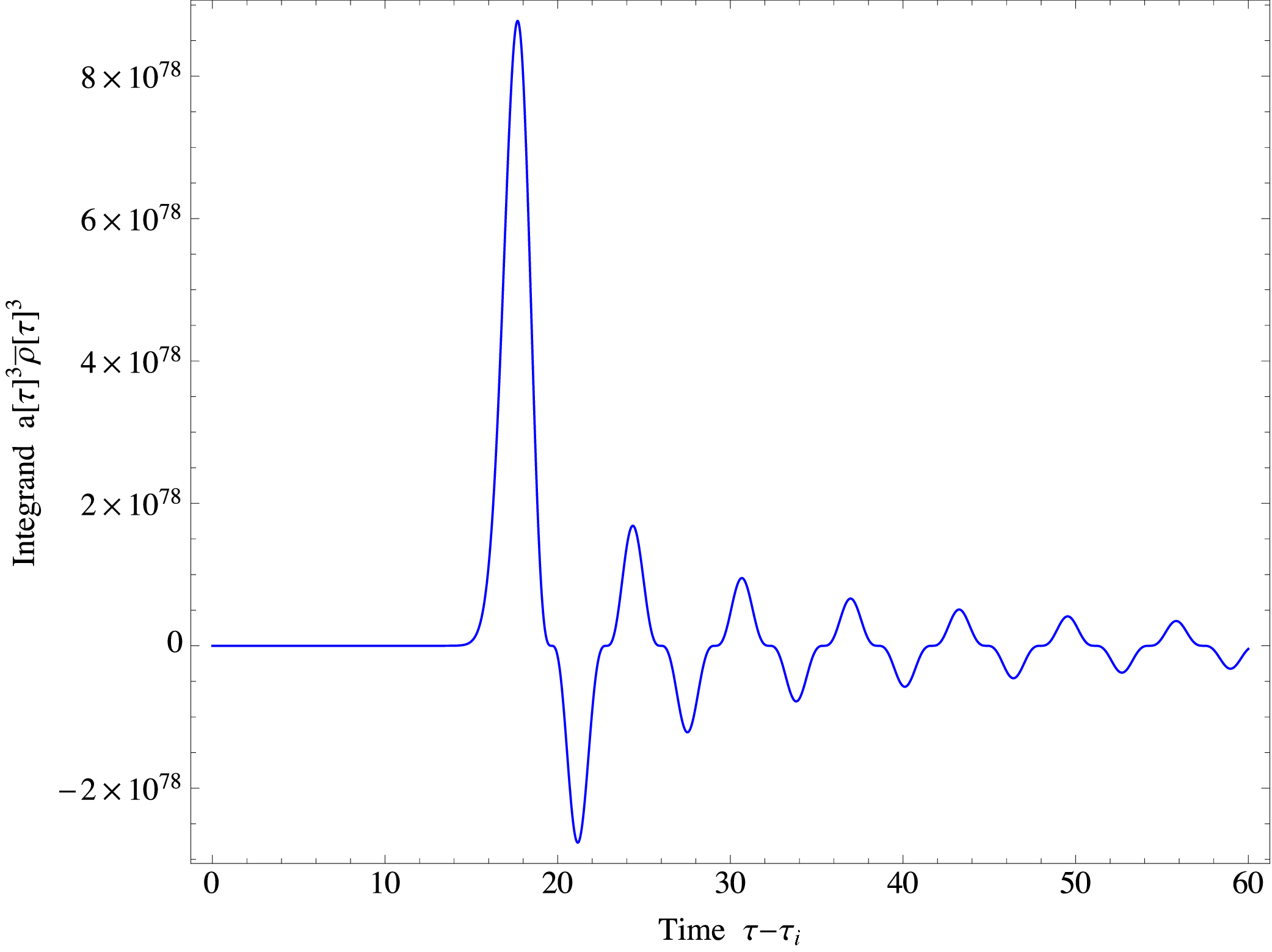}}
\caption{The integrand that appears in the asymmetry parameter $\as$ (see eqs.~(\ref{Aexp},\,\ref{fdef},\,\ref{barI1}), which is proportional to the total number of $\phi-\bar{\phi}$ particles. In this plot we have $n=3$ and initial conditions $\bar{\rho}_i=2\sqrt{60}$, $a_i=1$.
The large peak is in the latter phase of inflation; so this is where most of the $\phi$ (or anti-$\phi$) particles are produced.}
\label{Integrand}\end{figure}
This is seen in Fig.~\ref{Integrand} where we plot the integrand of $\bar{I}$ for $n=3$ as a function of $\tau$. So most of the $\phi$ (or anti-$\phi$) particles are generated in the latter stage of inflation.

It is useful to note that for quadratic inflation an approximate time for the start of the matter era is given by
$\tau-\tau_i \sim \sqrt{3/2}\,\bar{\rho}_i-1$, as will be derived in the next subsection.
Let us now compare to Fig.~\ref{Integrand}. In the figure we chose $\bar{\rho}_i=2\sqrt{60}$ (corresponding to $N_e\approx60$), which gives $\tau-\tau_i\approx 18$. Indeed this corresponds to the end of the sharp rise and fall of the integrand; it then begins to oscillate and redshift away in the matter era. This region is shifted to slightly earlier times in the inflationary era as we increase $n$, which we elaborate on in the next subsection.

Moreover, the denominator of $f_n$ is easily shown to be independent of $\tau_f$ at late times. In summary this means that $f_n$ approaches a constant that we denote $c_n$
that is independent of the initial and final times in this limit
\beq
c_n=f_n(\tau_i\to -\infty,\tau_f\to\infty). 
\label{cdef}\eeq
So our leading order approximation for the asymmetry takes on the simple form
\beq
\as = -c_n{\bp\mpl^{n-2}\over m^2}\sin(n\,\theta_i).
\label{Rapprox}\eeq

\begin{figure}
\center{\includegraphics[width=\columnwidth]{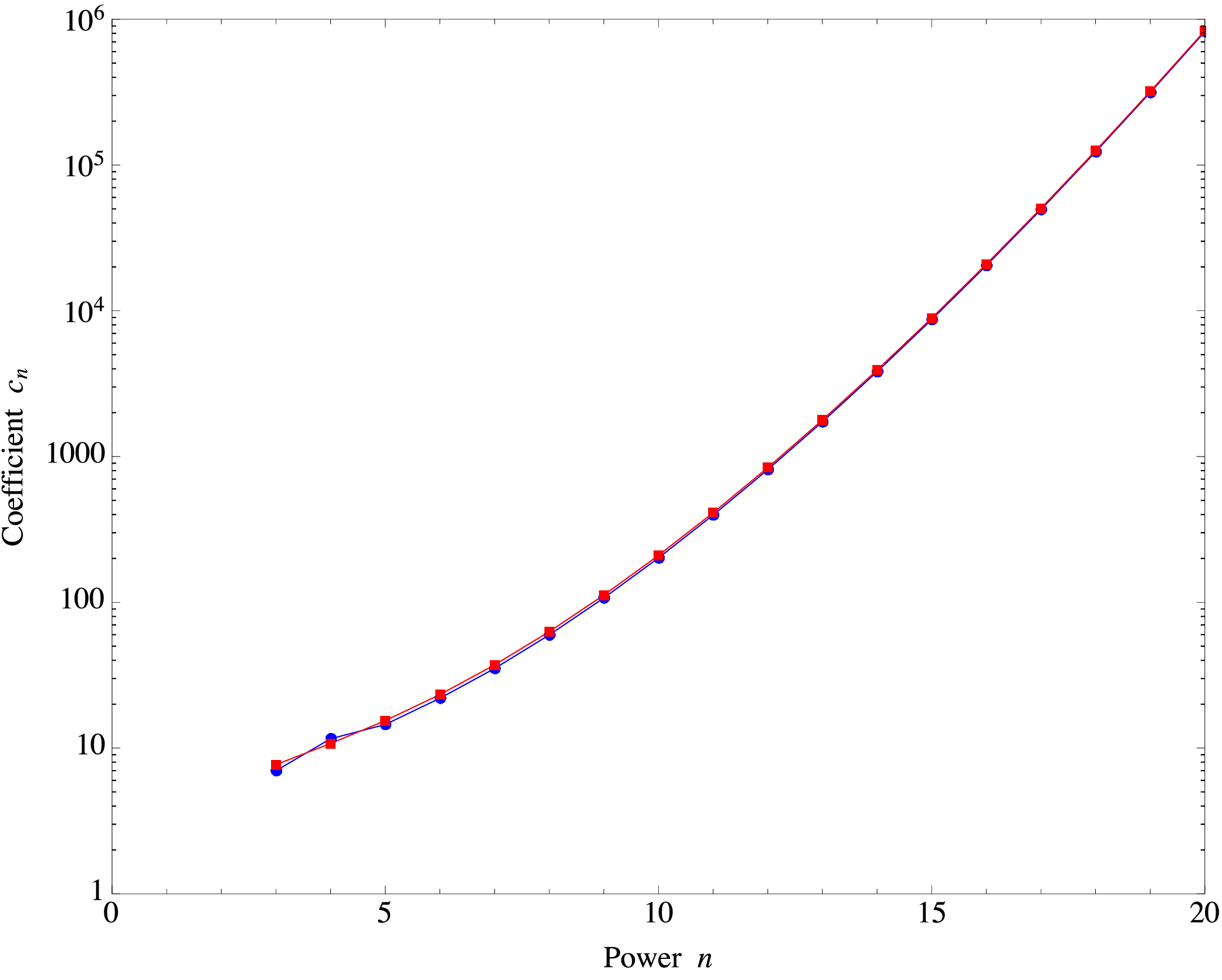}}
\caption{A plot of the asymmetry coefficient $c_n$ as a function of $n$. The blue is the exact numerical result and the red is the approximate analytical result given in eq.~(\ref{cappx2}) and derived in Section \ref{AppEstimate}.}
\label{Coefficient}\end{figure}

Numerically solving the dimensionless ordinary differential equation for $\bar{\rho}$ and then integrating, leads to the following results for the coefficient $c_n$ for the first few $n$
\bea
&&c_3\approx 7.0,\,\,\,c_4\approx 11.5,\,\,\,c_5\approx 14.4,\,\,\,c_6\approx 21.8,\nonumber\\
&&c_7\approx 34.8,\,\,\,c_8\approx59.3,\,\,\,c_9\approx107,\,\,\,c_{10}\approx 201.\,\,\,\,
\label{cnvalues}\eea
In the next subsection we derive an analytical expression for the coefficients $c_n$, valid in the large $n$ regime; see eq.~(\ref{cappx2}).
In Fig.~\ref{Coefficient}
we plot the coefficients $c_n$ for a range of $n\geq3$, comparing the numerical results to the upcoming approximate analytical results. 

\begin{figure}
\center{\includegraphics[width=\columnwidth]{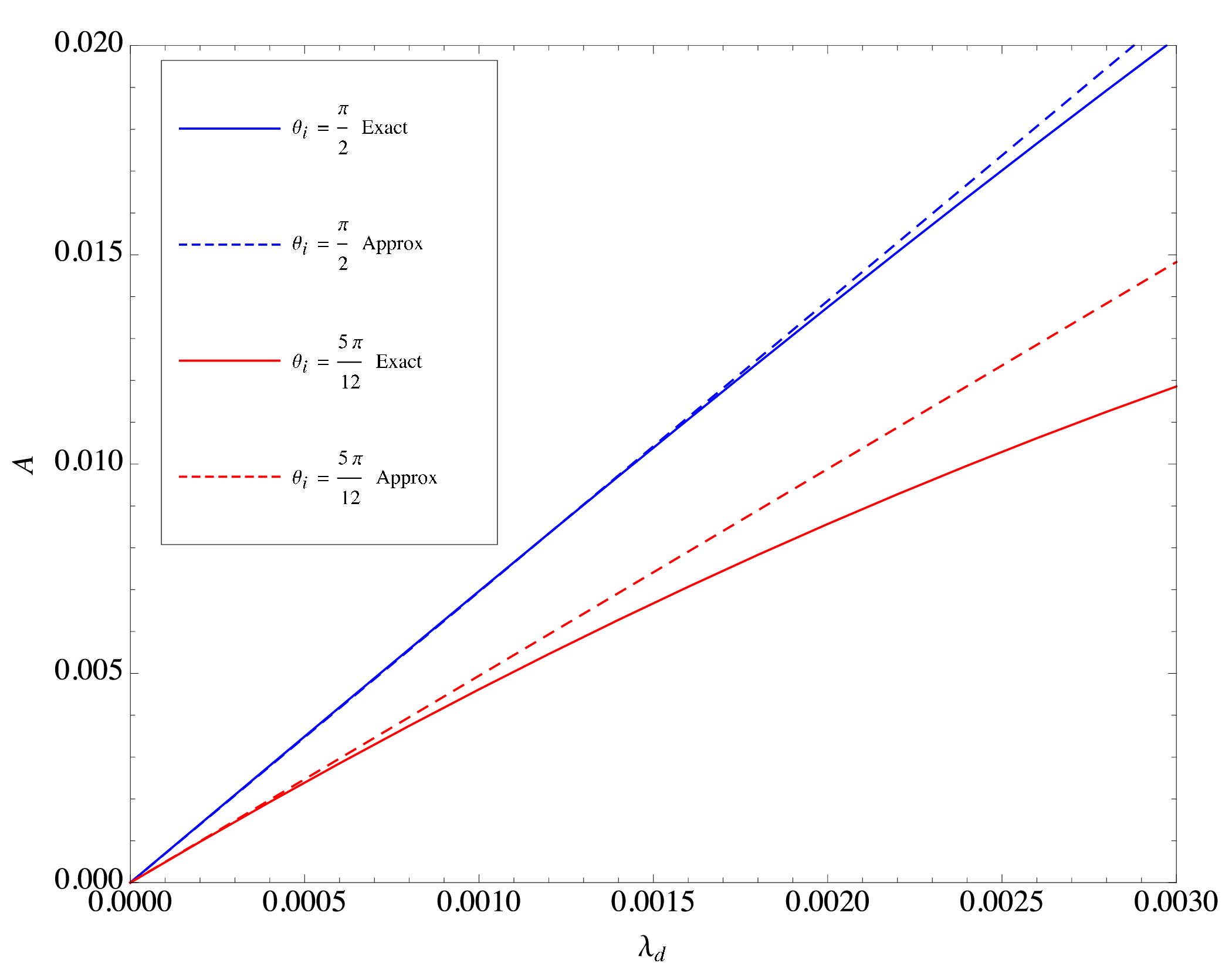}}
\caption{A plot of $\as$ versus $\bp_d\equiv\bp\mpl/m^2$ for $n=3$ with an initial amplitude during inflation of $\rho_i=2\sqrt{60}\,\mpl$. 
The solid curves are the exact numerical results and the dashed curves are the small $\bp$ analytical results given by eq.~(\ref{Rapprox}) with $c_3=7.0$.
The blue (upper) curves are for an initial angle of $\theta_i=\pi/2$ and the red (lower) curves are for an initial angle of $\theta_i=5\,\pi/12$.}
\label{RvsGamma}\end{figure}

We have also solved the full coupled complex system numerically; see Fig.~\ref{FieldEvolution} for a representative plot of the field evolution. 
In Fig.~\ref{RvsGamma}
we plot the full numerical result for $\as$ as a function of $\bp_d\equiv\bp\mpl/m^2$ for $n=3$ and we compare it to this leading order approximation. We have taken the field to begin during slow-roll inflation at $\rho_i=2\sqrt{60}\,\mpl$ (corresponding to $N_e\approx60$), which is relevant for the full numerical solution, though it is irrelevant to the leading order approximation as we discussed earlier.

In Fig.~\ref{RvsGamma} we have presented the results of two different initial angles $\theta_i$.
We note that the $\theta_i=\pi/2$ case is considerably more accurate than the $\theta_i=5\,\pi/12$ case. We can understand this difference as follows: In addition to the leading $\bp$ approximation, we expect higher order corrections. These higher order corrections should involve higher harmonics in $3\,\theta_i$. So while the first term is proportional to $\sin(3\,\theta_i)$, higher terms will include pieces proportional to $\sin(6\,\theta_i)$, etc. For special values of $\theta_i$, such as $\theta_i=\pi/2$, this piece proportional to $\sin(6\,\theta_i)$ vanishes, while it does not vanish for a generic value of $\theta_i$, such as $\theta_i=5\,\pi/12$. Furthermore, higher order terms will involve higher powers of $\bp$; this explains the departure from the analytical result as we increase $\bp$.

\subsection{Analytical Estimate}\label{AppEstimate}

In the previous subsection we reported on our numerical results for the coefficient $c_n$ in eq.~(\ref{cnvalues}) for $3\leq n\leq10$. Later, in Section \ref{High} we discuss an interesting possibility that high $n$ may be of interest. So here we would like to calculate the $c_n$ for high $n$. We will find that our result is surprisingly accurate even for small $n$.

In order to compute the asymmetry, we need $a(\tau)$ and $\bar{\rho}(\tau)$, which come from solving the dimensionless versions of eqs.~(\ref{rhosimp},\,\ref{Hsimp}).
In the slow-roll regime, the $\ddot{\bar{\rho}}$ is small in the equation of motion for $\bar{\rho}$ and the kinetic term $\dot{\bar{\rho}}^2/2$ is small in the equation for $\bar{H}$. To make this explicit, we rewrite these equations with factors of $\delta$ next to these small terms, i.e.,
\bea
&&\delta\ddot{\bar{\rho}}+3\bar{H}\bar{\rho}+\bar{\rho}=0,\\
&&\bar{H}^2={1\over3}\left({\delta\over2}\left(d\bar{\rho}\over d\tau\right)^{\!2}\!+{1\over2}\bar{\rho}^2\right).
\eea
This allows for a consistent power series expansion as follows
\bea
\bar{\rho}\amp=\amp\bar{\rho}_a+\delta\bar{\rho}_b+\ldots\\
\bar{H}\amp=\amp\bar{H}_a+\delta\bar{H}_b+\ldots
\eea
In the final result we will replace $\delta\to1$. For an approximation to $\bar{I}$, we only need the first few terms in the expansion. 
After a straightforward calculation, we find the solution
\bea
\bar{\rho}_a(\tau) \amp = \amp \bar{\rho}_i-\sqrt{2\over3}\,(\tau-\tau_i),\\
\bar{\rho}_b(\tau) \amp = \amp {1\over3}\left({1\over\bar{\rho}_a(\tau)}-{1\over\bar{\rho}_i}\right),\\
\bar{H}_a(\tau)\amp = \amp {\bar{\rho}_a(\tau)\over\sqrt{6}},\\
\bar{H}_b(\tau)\amp =  \amp {1\over3\sqrt{6}}\left({2\over\bar{\rho}_a(\tau)}-{1\over{\bar{\rho}}_i} \right).
\eea
Recall that the integrand of $\bar{I}$ is $\sim \bar{a}(\tau)^3\bar{\rho}(\tau)^n$. Since it is only a power law in $\bar{\rho}$, the first term $\bar{\rho}_a$ will suffice, while for $\bar{a}$ the first and second terms in $\bar{H}$ are required, as the scale factor is an exponential of this, so higher accuracy is required.
After integrating the Hubble parameter $\bar{H}_a+\bar{H}_b$, we find the following expression for the scale factor in the slow-roll regime, including a subleading correction
\beq
a(\tau) \approx {a_i\,\exp\!\left({\tau-\tau_i\over\sqrt{6}}\left(\bar{\rho}_i-{1\over3\bar{\rho}_i}\right)-{(\tau-\tau_i)^2\over6}\right)
\over\left(1-\sqrt{2\over3}{\tau-\tau_i\over\bar{\rho}_i}\right)^{\!1/3}}.
\eeq

For large $n$ the integral $\bar{I}(\tau_i,\tau_f)$ (eq.~(\ref{barI})) is dominated by the late stage of inflation rather than the matter era. For a concrete approximation lets use the slow-roll regime from the start of inflation until the slow-roll parameter reaches $\epsilon=3$, which corresponds to the equation of state $w=0$; the start of the matter era. This corresponds to using an endpoint of integration of
\beq
\tau_f=\tau_i+\sqrt{3\over2}\,\bar{\rho}_i-1
\eeq
and ignoring the contribution from the matter era as it is relatively small for large $n$. 

We now perform a change of variable from $\tau$ to $b$ where 
\beq
b\equiv \sqrt{3\over2}\,\bar{\rho}_i-(\tau-\tau_i).
\eeq
The corresponding approximate value of the integral becomes
\bea
\bar{I} &\approx& a_i^3\left(2\over3\right)^{(n-1)/2}{\bar{\rho}_i\,e^{3\bar{\rho}_i^2/4}\over\sqrt{e}}\nonumber\\
\amp\amp\times \int_1^{\sqrt{3\over2}\bar{\rho}_i} db\,b^{n-1}\exp\left(-{b^2\over2}+{b\over\sqrt{6}\,\bar{\rho}_i}\right).
\eea
Since we are assuming the universe began deep in the slow-roll regime, we can take $\bar{\rho}_i$ large and ignore the $b/\bar{\rho}_i$ term inside the exponential and send the upper limit of the integral to $\infty$. 
We are then able to rewrite the integral as follows
\beq
\bar{I}\approx a_i^32^{n-3/2}3^{-(n-1)/2}{\bar{\rho}_i\,e^{3\bar{\rho}_i^2/4}\over\sqrt{e}}\Gamma_{1\over2}(n/2),
\label{barI}\eeq
where $\Gamma_a$ is the incomplete gamma function, defined as
\beq
\Gamma_{a}(z)\equiv \int_a^\infty dt\, t^{z-1}\,e^{-t}.
\eeq
We now form the coefficient $c_n$ defined by eqs.~(\ref{fdef},\,\ref{cdef}). To do so, we need to divide $\bar{I}$ by $a(\tau)^3\bar{\varepsilon}(\tau)$. This cancels the factor of $a_i^3\bar{\rho}_i\exp\!\left(3\bar{\rho}_i^2/4\right)$ in (\ref{barI}), as this is just an extensive factor and independent of $n$. The final result for $c_n$ is then
\beq
c_n\approx \tilde{c}\,2^{n/2}3^{-n/2}\,n \,\Gamma_{1\over2}(n/2),
\label{cappx2}\eeq
where $\tilde{c}$ absorbs any $\mathcal{O}(1)$ factors that are independent of $n$; it accounts for the transition of $\bar{\rho}$, $a$ from inflation to matter eras, and we find its value to be $\tilde{c}\approx 6.64$. 
This result for $c_n$ is asymptotically correct for large values of $n$ and also provides a good estimate for moderate values of $n$; see Fig.~\ref{Coefficient} for comparison to the exact numerical results.

\section{Decay into Baryons}\label{Decay}

We would now like to compute the relationship between this asymmetry parameter $\as$ defined in eq.~(\ref{asdef}) and the final productions of baryons.
Recall that the baryon asymmetry is defined as the ratio of baryon difference to photon number at late times
\beq
\eta={(\nb-\nbb)_f\over (n_\gamma)_f}={(\Nb-\Nbb)_f\over(N_\gamma)_f},
\eeq
where $f$ indicates the late time, or ``final" value, after decay and thermalization.
We would like to relate this to the $\phi$ asymmetry parameter $\as$ and the details of how $\phi$ decays. 

We associate with each $\phi$ particle a baryon number $\bphi$; for instance $\bphi=1$ or $\bphi=1/3$ in simple models (see Section \ref{Particle}). We assume that the decay of $\phi$ and all subsequent interactions is baryon number conserving, so we can relate the final number to the initial number as follows
\beq
(\Nb-\Nbb)_f=\bphi(\Nphi-\Nphib)_i,
\eeq
where $i$ indicates the early time, or ``initial" value, before decay and thermalization (but well after the baryon violating processes have stopped).     We can now re-write $\eta$ as
\bea
\eta \amp =\amp  \bphi {(\Nphi-\Nphib)_i\over(N_\gamma)_f} = \bphi \as{(\Nphi+\Nphib)_i\over(N_\gamma)_f}.
\eea
So we need to evaluate the number of $\phi$ particles at early times and the number of photons at late times.

At early times we can relate the number of $\phi$ particles to the energy density as follows
\beq
(\Nphi+\Nphib)_i=V_{com}(a^3 \varepsilon)_i/m
\eeq
and the energy density is related to the Hubble parameter as
\beq
(\varepsilon)_i=3\mpl^2 (H^2)_i.
\eeq

On the other hand, at late times we can relate the number of photons to the number density as follows
\beq
(N_\gamma)_f = V_{com}(a^3 n_\gamma)_f
\eeq
and the number density is related to the temperature as follows
\beq
(n_\gamma)_f={2\zeta(3)\over\pi^2}(T^3)_f.
\eeq

Hence we express $\eta$ as
\beq
\eta = {3\,\pi^2\bphi\over 2\,\zeta(3)}{\mpl^2 \as\over m}{(a^3 H^2)_i\over(a^3 T^3)_f}.
\label{etarewrite}
\eeq
Note that both the numerator and denominator here are separately time independent. 
It is non-trivial to exactly compute these final parameters as a function of initial conditions as it depends on the details of the decay and thermalization. However, a good approximation arises by assuming that the thermalization is rapid. This means that we can simply evaluate both the ``initial" and ``final" quantities around the time of decay. 

We denote the decay rate of the $\phi$ field as $\dr$. 
Thermalization occurs around $H\approx \dr$ \cite{Kofman:1997yn}. The energy density in the radiation era is
\beq
\varepsilon = {\pi^2\over 30}\,g_*T^4,
\eeq
where $g_*$ is the number of relativistic degrees of freedom, which is typically $g_*\sim10^2$.
Then by setting $\varepsilon = 3 H^2\mpl^2\approx 3\dr^2\mpl^2$ and solving for $T$, we have an estimate of the reheat temperature
\beq
T_r \approx\left(90\over g_*\pi^2\right)^{\!1/4}\!\dr^{1/2} \mpl^{1/2}.
\label{retemp}\eeq
Substituting $T$ with the reheat temperature $T_r$ and $H\approx\dr$ into (\ref{etarewrite}), we obtain
\beq
\eta\approx {\fud\,\pi^{7/2}g_*^{3/4}\bphi\over 2^{7/4}3^{1/2}5^{3/4}}{\as\,\dr^{1/2}\mpl^{1/2}\over m},
\eeq
where $\fud$ is an $\mathcal{O}(1)$ fudge factor that accounts for the details of the transition from the $\phi$ era to the thermal era. The precise value of $\fud$ is not important for our qualitative conclusions, but we do expect it to be of order $1$.

Finally, we insert the expression for $\as$ from eq.~(\ref{Rapprox}) to obtain our result for the baryon-to-photon ratio
\beq
\eta\approx -c_n{\fud\,\pi^{7/2}g_*^{3/4}\bphi\over 2^{7/4}3^{1/2}5^{3/4}}{\bp\,\dr^{1/2}\mpl^{n-3/2}\over m^3}\sin(n\,\theta_i).
\label{etaexp}\eeq
In order for this $\eta$ to match with the observed value $\eta_{obs}\approx 6\times 10^{-10}$, we require $\bp\,\dr^{1/2}\mpl^{n-3/2}/m^3$ to take on a particular value, namely
\bea
{\bp\,\dr^{1/2}\mpl^{n-3/2}\over m^3}\Bigg{|}_{\mbox{\tiny{req}}} \amp \approx\amp 7\times 10^{-11}c_n^{-1}\nonumber\\
\amp\times\amp  \Bigg{(} {g_*^{3/4}\over 30}\fud\, \bphi|\sin(n\,\theta_i)|\Bigg{)}^{\!-1},
\label{Gammacons}
\eea
where the subscript ``req" indicates that this is the ``required" value for agreement with $\eta_{obs}$.
Inside the parenthesis is a term that should be $\mathcal{O}(1)$, since we expect 
\beq
|\sin(n\,\theta_i)|\sim 1,\, \bphi\sim 1,\, \fud\sim 1,\, g_*\sim10^2,
\eeq 
(provided $\theta_i$ is not near a special value, as mentioned earlier). As a result, the prefactor of $\sim 7\times 10^{-11}c_n^{-1}$ is most important here.

\section{Constraints from Inflation}\label{Inflation}

\subsection{Quadratic Inflation}

During inflation we assume that the potential is dominated by the symmetric $m^2\phi^*\phi$ term, and so to first approximation the motion is radial. We can thus keep the phase of our complex field fixed: $\phi=\rho \, e^{i\theta_i}/\sqrt{2}$, and write the potential and the so-called slow-roll parameters as
\bea
V&=&{1\over2}m^2\rho^2,\\
\epsilon_{sr} &\equiv& {\mpl^2\over2}\left(V'\over V\right)^{\!2} = {2\mpl^2\over\rho^2},\\
\eta_{sr} &\equiv& \mpl^2{V''\over V} = {2\mpl^2\over\rho^2}.
\eea
Also, the number of e-foldings is given by
\beq
N_e={1\over \mpl}\int^{\rho_i}_{\rho_{end}} \!{d\rho\over\sqrt{2\epsilon}}={\rho_i^{2}-\rho_{end}^2\over 4\mpl^2}.
\eeq
Hence the modes that describe the universe on large scales are emitted when the field value is
\beq
\rho_i \approx 2\sqrt{N_e}\,\mpl
\eeq
and inflation ends for $\rho_{end}\sim\mpl$ (when $\epsilon_{sr}\sim1$).

For any simple single field slow-roll model of inflation, the squared amplitude of density fluctuations is predicted to be
\beq
\Delta_R^2 = {V_i\over 24\pi^2\mpl^4\epsilon_{sr}} \label{DeltaR},
\eeq
where $V_i$ is the potential energy when modes leave the horizon.
The observed value is $\Delta^2_{R,obs}\approx 2.45\times 10^{-9}$ from WMAP and Planck data \cite{Hinshaw:2012aka,Ade:2013uln}.
For the simple quadratic model, we have
\beq
\Delta_R^2\approx {N_e^2\,m^2\over6\,\pi^2\mpl^2}.
\label{DeltaR2}
\eeq
Hence in order to have the correct amplitude of density fluctuations, the mass of the field must be
\bea
m\amp\approx\amp {\sqrt{6}\,\pi\Delta_{R,obs}\mpl\over N_e}
\approx1.5\times10^{13}\,\mbox{GeV}\left(60\over N_e\right).\,\,\,
\eea

We have indicated that $N_e=60$ is a good reference value, but its precise value can be smaller if the reheat temperature is low. The precise relationship between the number of e-foldings of inflation and the scale of interest $k$
is given by \cite{LiddleBook}
\bea
N_e\amp=\amp62-\ln{k\over a\,H}+{1\over4}\ln{V_i\over(10^{16}\,\mbox{GeV})^4}\nonumber\\
\amp\amp+{1\over4}\ln{V_i\over V_{end}}-{1\over12}\ln{V_{end}\over \varepsilon_{reh}}
\eea
where $V_{end}$ is the potential energy at the end of inflation, and $\varepsilon_{reh}$ is the energy density at the start of reheating. For typical high scale inflation models, such as quadratic inflation, $V_i\sim(10^{16}\,\mbox{GeV})^4$. For efficient reheating this leads to $N_e\sim 60$, while for inefficient reheating, say $H_{reh}\sim 1$\,eV, we have $N_e\sim 50$, for a typical CMB scale $k$.
For $N_e\sim 50-60$ the spectral index $n_s=1-6\epsilon_{sr}+2\eta_{sr}\approx1-2/N_e$
is in excellent agreement with observation $n_s\approx 0.96$ \cite{Hinshaw:2012aka,Ade:2013uln}.

On the other hand a small problem with these quadratic models is the tensor-to-scalar ratio, which is
$r=16\,\epsilon_{sr}\approx 8/N_e$ which is somewhat large and just outside the 2 sigma region of recent Planck CMB constraints \cite{Ade:2013uln}. However, it is important to note that the details of the inflationary potential are not crucial to our baryogenesis mechanism. We can easily construct other potentials which ``flatten" at large field values, decreasing the tensor-to-scalar ratio, and still provide the same qualitative mechanism for baryogenesis as outlined in this work. For the present purposes it is enough to use the above mass scale $m$ and number of e-foldings $N_e$ as characteristic values (with $m$ having the greatest variability between different inflationary models). 

\subsection{Implications for Baryogenesis}\label{Implications}

An important constraint is that the symmetry breaking term in the potential $\bp(\phi^n+\phi^{*n})$ be subdominant during inflation. Since this contribution to the potential goes negative at large field values, we obviously need it to be small during inflation. Writing $\phi$ in terms of polar variables, the constraint is
\beq
{\bp\over 2^{n/2-1}}\rho_i^n\cos(n\,\theta_i)\ll {1\over2}m^2\rho_i^2.
\eeq
Then using $\rho_i=2\sqrt{N_e}\,\mpl$, and demanding this to be true for all $\theta_i$, gives the following upper bound on $\bp$
\beq
\bp\ll\bp_0\equiv{m^2\over 2^{n/2}N_e^{n/2-1}\mpl^{n-2}}.
\label{gammacons}\eeq
For typical values of mass and duration of inflation (e.g., $m\sim 10^{13}$\,GeV and $N_e\sim 60$), this provides an important bound on $\bp$ for each $n$. 

We now use the threshold value $\bp_0$ to rewrite the condition for the correct baryon-to-photon ratio (\ref{Gammacons}) as a condition on the decay rate
\bea
\drreq\amp\approx\amp 10^{-7}\,\mbox{eV}\times 2^{n+1}N_e^{n-2}c_n^{-2}   \left(\bp_0\over\bp\right) \nonumber\\
\amp\times\amp\left(m\over 10^{13}\,\mbox{GeV}\right)^{\!2} \Bigg{(} {g_*^{3/4}\over 30}\fud\, \bphi|\sin(n\,\theta_i)|\Bigg{)}^{\!-2}\!.\,\,\,\,
\label{drcond}\eea
To provide concrete quantitative results for the required decay rate, we need to choose some characteristic values for parameters.
For instance, if we assume that the coupling $\bp$ is a factor of 10 smaller than its inflationary upper bound $\bp_0$, and take $\fud \bphi|\sin(n\,\theta_i)|\approx 1$, $m\approx 1.5\times 10^{13}$\,GeV, $N_e\approx 55$, $g_*\approx 10^2$, and we insert the $c_n$ from eq.~(\ref{cnvalues}) (or (\ref{cappx2})), then for each of the different $n$ (from 3 to 10) we find that the required decay rate is
\bea
&&n=3\Rightarrow\drreq\approx 4\times 10^{-5}\,\mbox{eV},\nonumber\\
&&n=4\Rightarrow\drreq\approx 2\times 10^{-3}\,\mbox{eV},\nonumber\\
&&n=5\Rightarrow\drreq\approx 10^{-1}\,\mbox{eV},\nonumber\\
&&n=6\Rightarrow\drreq\approx6\,\mbox{eV},\nonumber\\
&&n=7\Rightarrow\drreq\approx2\times 10^2\,\mbox{eV},\nonumber\\
&&n=8\Rightarrow\drreq\approx 9\times 10^3\,\mbox{eV}\nonumber\\
&&n=9\Rightarrow\drreq\approx3\times 10^{5}\,\mbox{eV},\nonumber\\
&&n=10\Rightarrow\drreq\approx 10^7\,\mbox{eV}.
\label{drvalues}\eea
It is important to compute the associated reheat temperature using eq.~(\ref{retemp}).
For $n=3$, corresponding to the lowest value of $\drreq$, the reheat temperature is still substantial: $T_r\sim 10^2$\,GeV. 
For higher $n$, $\drreq$ increases, so too does $T_r$; e.g., for $n=10$, $T_r\sim 10^8$\,GeV. All much higher than $\sim$\,MeV; the characteristic temperature of the universe at the onset of big bang nucleosynthesis.

Having established a consistent cosmology of inflation, reheating, and big bang nucleosynthesis, we would now like to go further and explore if such decay rates are plausible in various particle physics frameworks.

\section{Particle Physics Models}\label{Particle}

In the following, we introduce two different types of particle physics models. First, we consider $\phi$ to be a gauge singlet, and second, we consider $\phi$ to carry color. For each model we estimate the decay rates and compare to the required decay rates $\drreq$ to obtain $\eta_{obs}$.

Our upcoming estimates of the decay rates will involve a perturbative analysis. At the end of inflation when the field value is large, a non-perturbative regime involving parametric resonance can sometimes take place. However its existence and/or effectiveness is highly model dependent. In particular there can be self-resonance and/or resonance into other fields. The possibility of self-resonance could arise from the nonlinear $\sim\lambda(\phi^n+\phi^{*n})$ term in the action. However, it can be shown that since this term is assumed to be sub-dominant during inflation, it is even more sub-dominant after inflation, and the corresponding resonance is entirely inefficient and negligible. So this justifies our perturbative treatment while ignoring non-perturbative self-resonance. The possibility of resonance into other fields is interesting and is possible future work that we discuss later in the discussion section. In this case its existence depends on the value of the couplings to other fields, such as the Higgs. It is radiatively stable and self consistent to assume these couplings are sufficiently small that this too is ignorable. We assume, for simplicity, that this is the case in the present work.

\subsection{High Dimension Operators}\label{High}

In the simplest case one can take $\phi$ to be a gauge singlet. 
In this case there are many operators which could couple $\phi$ to Standard Model degrees of freedom. This is problematic, because there is no argument based on degree of freedom counting (namely, gauge redundancy) as to why the particle number contained in $\phi$ should only decay into quarks, etc, generating baryons. For instance, one could include operators, such as $\sim\phi\, H^\dagger H$ or $\phi\bar{f}f$, which would violate baryon number conservation and $\phi$ would just decay to Higgs or fermions, etc. 

However, a natural way around this problem of $\phi$ decaying into non-baryonic particles is to suppose that the global $U(1)$ symmetry is almost an exact symmetry of nature (or at least in the $\phi$ sector). This assumption is consistent with the requirement that the $U(1)$ breaking is small for a consistent model of inflation, which is technically natural. Although the symmetry is not expected to be exact.

The reason for this is the following: We begin by imaging we have some $U(1)$ global symmetry. However, we know that global symmetries cannot be exact in quantum gravity (one way to see this is from the ``black-hole-no-hair-theorem" which forbids a black hole to carry any baryon number, and therefore the $U(1)$ baryon symmetry cannot be exact). So a complete theory should allow for at least a weak breaking of the symmetry. Symmetries are often broken by some high dimension operator, representing the breaking due to some microscopic, perhaps Planckian, physics. Another way to argue this is to just impose a $\mathbb{Z}_n$ symmetry and assume $n$ is greater than 4, which is a reasonable model building assumption. 
For high $n$, the breaking parameter will need to satisfy $\bp\lesssim(\mbox{few}/\sqrt{G})^{4-n}$ to be be consistent with inflation. This is compatible with quantum gravity expectations.

This suggests the intriguing possibility of using a high value of $n$ in the breaking term. For instance, we can imagine that the $U(1)$ symmetry breaking occurs at dimension $n\ge8$ operators. If this is the case, then all low dimension operators, such as $\sim\phi\, H^\dagger H$, that break the $U(1)$ or discrete $\mathbb{Z}_n$ symmetries, would be forbidden. 

For definiteness, let us take $n=8$ and suppose that the $U(1)$ symmetry is intact for all dimension 7 or lower (which is protected against radiative corrections by a $\mathbb{Z}_8$ symmetry). Since $\phi$ carries baryon number, then up to dimension 7 it could only decay into quarks. It is well known that at dimension 6 the Standard Model degrees of freedom allow for gauge singlet operators carrying baryon number of the form $q\,q\,q\,l$. 
In fact there are 5 types of such operators \cite{Grzadkowski:2010es}, but the full details do not concern us here.
In order to construct a $U(1)$ invariant operator, we multiply the $q\,q\,q\,l$ operator by $\phi$ ($\phi^*)$, and build the following $U(1)$ symmetric dimension 7 operator
\beq
\Delta\mathcal{L}\sim{c\over \Lambda^3}\phi^*q\,q\,q\,l+\mbox{h.c},
\eeq
where we are suppressing indices for brevity.
Here we have introduced an energy scale $\Lambda$ that sets the scale of new physics (and the cutoff on the field theory) and $c$ is some dimensionless coupling.
In this case we have $\bphi=1$, as this operator causes $\phi$ to decay into 3 quarks (and a lepton).

The decay rate associated with this operator is roughly
\beq
\dr(\phi\to q+q+q+l)\sim {c^2\over 8\pi} {m^7\over \Lambda^6}.
\eeq
We now compare this to the required decay rates from (\ref{drvalues}).
For the case of $n=8$, we would need this to be $\dr\sim 10^4$\,eV.
For $m\approx 1.5\times 10^{13}$\,GeV and $c=\mathcal{O}(1)$, we find that the model has the required decay rate for
\beq
\Lambda\sim 10^{16}\,\mbox{GeV},
\eeq
which is intriguingly around the GUT scale.
Similarly, if we push the breaking off to even higher operators, such as $n=10$ or $n=12$, then the model requires slightly smaller $\Lambda$, approaching $\Lambda\sim 10^{15}$\,GeV, which is still close to the GUT scale. 

Since we are utilizing high dimension operators we are studying only an effective field theory. As such, we need to check that the parameters are within the regime of validity of the theory. There are two basic constraints: The first is that the physics that is being integrated out is below the scale of quantum gravity $E_{QG}$, i.e., it would be rather unrealistic if we were appealing to physics beyond the quantum gravity scale to UV complete the theory. The second is that the inflationary regime, which probes scales of order Hubble, is below the cutoff of the field theory. Together these two constraints imply
\beq
H_i\ll\Lambda\ll E_{QG}.
\eeq
Since $H_i\sim 10^{13,14}$\,GeV for high scale inflation models, and $E_{QG}\sim 10^{18,19}$\,GeV, then we require values for $\Lambda$  around $\Lambda\sim 10^{15,16}$\,GeV for the effective field theory to make sense. Remarkably, this is precisely the value found above to obtain $\eta_{obs}$.

To summarize, if the visible sector has an almost exact $U(1)$ symmetry that is broken only by high dimension operators ($n\ge 8$), then the leading operator, allowed by symmetry, that mediates $\phi$ decay is dimension 7 and the decay products are quarks and leptons. After imposing constraints from inflation on the breaking parameter and by taking the dimension 7 operator to be controlled by the GUT scale, which is within the regime of validity of the effective field theory, we can obtain the observed matter/anti-matter asymmetry $\eta_{obs}$.

\subsection{Colored Inflaton}\label{Colored}

Another possibility is to allow the inflaton to carry color and to transform under the fundamental representation of $SU(3)_c$. 
We give $\phi$ a color index, $i=r,w,b$, and allow for ``up" $\phi_u$ and ``down" $\phi_d$ versions and different generations labelled by $g$.
In addition to the usual symmetric mass term, etc, we can construct $U(1)$ violating terms in the potential that respect the $SU(3)_c$ symmetry. For instance, at dimension $n=3$, we can introduce the breaking term
\beq
V_b(\phi,\phi^*) = \bp^{gg'\!g''}  \varepsilon_{ii'\!i''} \, \phi_{ug}^{i}\,\phi_{dg'}^{i'} \, \phi_{dg''}^{i''}+\mbox{h.c},
\eeq
where $\varepsilon_{ijk}$ is the totally anti-symmetric tensor, and we have summed over color indices and different generations. 
We need one ``up" and two ``down" for a gauge singlet under electric charge. We also need at least 2 different generations, or otherwise the anti-symmetric tensor will cause the term to vanish.
For simplicity we have written down the leading $U(1)$ violating operator; dimension $n=3$. As in the previous subsection we could imagine that the symmetry is only broken by some higher dimension operators; such a generalization is straightforward.

In order to compute the generated particle number, we should repeat the analysis from Sections \ref{Model}, \ref{Results}, but now with various indices to track. We expect the basic qualitative conclusions from those sections to be the same, so we will not go through those details here. 
Moreover, corrections from gluons should be small at these high energies due to asymptotic freedom \cite{GrossWilczek,Politzer}.

Since $\phi$ carries color, we can readily build operators that mediate $\phi$ decay into quarks, while respecting the global symmetry, such as the following dimension 4 operator
\beq
\Delta\mathcal{L}\sim y\,\phi^{i*} q^i \bar{f}+\mbox{h.c},
\eeq
where $f$ is some color neutral fermion and $y$ is a type of Yukawa coupling. In this case we have $\bphi=1/3$, as this operator causes $\phi$ to decay into 1 quark (and another fermion). This decay rate is roughly
\beq
\dr(\phi\to q+\bar{f})\sim{y^2\over 8\pi}m.
\eeq
For high scale inflation, such as quadratic inflation that we discussed earlier, the inflaton mass is large $m\sim 10^{13}$\,GeV, so one would require an extremely small value of $y$ to obtain decay rates comparable to the required values we computed earlier in eq.~(\ref{drvalues}) (of course that was only for the gauge singlet case, but we expect similar values within an order of magnitude). For example, in order to have $\dr\sim$\,eV, we would need $y\sim 10^{-10}$. In generic models of particle physics, such a small dimensionless coupling would not normally be radiatively stable. In certain settings, such as supersymmetry (which would provide extra motivation for the existence of such colored scalars, or ``squarks"), one could examine if some non-renormalization theorem may help to stabilize $y$ at such small values. 

Perhaps a more promising possibility would be to push the $U(1)$ symmetry breaking parameter to values much smaller than is required by the inflationary constraint. This is, as mentioned earlier, technically natural. Recall from eq.~(\ref{drcond}) that the required decay rate to obtain the correct value of $\eta$ is inversely proportional to $\bp$. So by making $\bp$ extremely small, we can raise the values in eq.~(\ref{drvalues}) to much higher values (recall those values were based on $\bp\sim\bp_0/10$). A consistent model would then only require a moderately small value for $y$.

\section{Observational Consequences}\label{Observational}

\subsection{Large Scale Dipole}

As in the original Affleck-Dine model, an interesting feature of this mechanism is that the dynamics respect C (and CP) symmetry, but it is broken spontaneously by the state of the universe. In the present work, this is due to the initial value of the inflaton field in the complex plane. It is often thought that any initial conditions are wiped out by inflation, but the vev of the inflaton is evidently not. Instead this initial condition has a very important consequence for the final matter/anti-matter in the universe; it determines if the imbalance favors matter or anti-matter. Recall from eq.~(\ref{etaexp}) that $\eta$ depends on the initial angle of the inflaton
\beq
\eta\propto -\sin(n\,\theta_i).
\eeq
Since we expect the initial angle to be randomly distributed in the domain $\theta_i\in[0,2\pi]$, then we expect $\eta$ to be a random variable, with 50\% chance positive (matter domination; see the blue curve in Fig.~\ref{FieldEvolution}) and 50\% chance negative (anti-matter domination; see the red curve in Fig.~\ref{FieldEvolution}).

If inflation only lasted for the minimum number of e-foldings, say, 50 -- 60, then the initial $\theta_i$ angle can be inhomogeneous on the scale of order today's horizon. This translates directly into an inhomogeneity in $\eta$ on the corresponding scale. 
Recent Planck results have bounded the curvature of the universe to be small \cite{Ade:2013uln}
\beq
|\Omega_k|\lesssim 10^{-2},
\eeq
which indicates that inflation did last at least slightly longer than the absolute minimum required. But there exist various arguments regarding fine tuning of the inflaton potential to suggest that it may not have lasted much more than this \cite{Freivogel:2005vv}. If this is the case, then a small spatial variation would arise in $\eta$. Since this would be a super-horizon mode, it would appear as a dipole across the universe. Its magnitude would be correlated with $\Omega_k$ and  potentially detectable. In fact it could conceivably be related to existing CMB anomalies \cite{Ade:2013nlj}.

Another possibility, is that our pocket universe arose from bubble nucleation in an eternally inflating false vacuum. In this case there exist arguments based on Coleman De Luccia \cite{Coleman:1980aw} tunneling to indicate that the resulting bubble would be completely homogeneous (plus small quantum fluctuations). This would forbid the formation of such a dipole. On the other hand, this process would produce an infinite number of different bubbles, each with a different value of $\theta_i$; a multiverse of different baryon densities. 

\subsection{Isocurvature Fluctuations}

Quantum fluctuations from inflation provide an excellent candidate for the origin of density fluctuations in the universe.
In simple single field models, only a curvature (``adiabatic") fluctuation is generated, due to fluctuations in the inflaton. Earlier, in eq.~(\ref{DeltaR}), we mentioned the amplitude of these fluctuations $\Delta_R^2$. 

For multi-field inflationary models, an isocurvature (``entropic") fluctuation is also generated \cite{Bartolo:2001rt}. This is due to quantum fluctuations in the field orthogonal to the classical field trajectory, which leaves the total density unchanged. Since we are studying complex (two field) inflation,  there will be isocurvature fluctuations in $\phi$, and these will generate isocurvature fluctuations in the baryon density.

The orthogonal fluctuations are in the angular variable $\theta$, with mean value $\theta_i$ and $\langle|\delta\phi|^2\rangle_{iso}=\rho_i^2\langle\delta\theta^2\rangle/2$. During inflation, this orthogonal fluctuation is a  light (Goldstone) field with an approximately scale invariant spectrum of fluctuations of amplitude \cite{Linde:2005dd}
\beq
\langle\delta\theta^2\rangle\approx{\gamf^2\over4\pi^2}{H_i^2\over \rho_i^2},
\label{deltatheta}\eeq
where $H_i$ is the Hubble parameter during inflation, $\rho_i$ is the corresponding radial field value, and $\gamf$ is an $\mathcal{O}(1)$ factor that accounts for Brownian motion during inflation. 

In the radiation era, after all annihilations have occurred, but well before equality, the baryon density is related to $\theta$ by $\nb\propto-\sin(n\,\theta)$. Working to first order, we can relate the early time baryon fluctuations to the $\theta$ fluctuations as
\bea
{\delta\nb\over\nb}\Big{|}_{early}= n\,\delta\theta\cot(n\,\theta_i).
\label{deltaeta}\eea
Note that near special values of the initial angle $\theta_i={p\pi\over n}\,|\,p\in\mathbb{Z}$, this leading order approximation breaks down. This first order approximation is valid whenever there is a non-negligible average baryon number, which is necessary for baryogenesis.

We would now like to determine the corresponding isocurvature fluctuations in the CMB temperature.
By definition, the isocurvature fluctuation satisfies
\beq
0=\delta\rho_{iso}=m_b\delta\nb+m_{cdm}\delta n_{cdm}+\delta\rho_\gamma,
\label{isodef}\eeq
where we have included baryons, dark matter, and radiation.
We assume that for all species $s$, other than baryons, the fluctuations are adiabatic, i.e., $\delta n_s/n_s-\delta n_\gamma/n_\gamma=0$ for $s\neq b$. 
If we divide eq.~(\ref{isodef}) by $\rho_b=m_b \nb$, it is easy to show that this implies $\delta\nb/\nb\gg\delta\rho_\gamma/\rho_\gamma(=4\,\delta T/T=(4/3)\delta n_\gamma/n_\gamma)$ at early times. Hence the entropy perturbation is
\beq
{\delta\eta\over\eta}={\delta\nb\over\nb}-{\delta n_\gamma\over n_\gamma}\approx {\delta\nb\over\nb}\Big{|}_{early}.
\label{deltaeta2}\eeq
Using this result and using the fact that entropy perturbations are approximately conserved outside the horizon, we find that the isocurvature contribution to the temperature at late time re-entry is
\beq
\left({\delta T\over T}\right)_{\!iso}= -{6\over15}{\Omega_b\over\Omega_m}{\delta\eta\over\eta}.
\label{deltaT}\eeq
We have included a factor of 6/5 due to the Sachs-Wolfe effect \cite{Sachs} (whose details we shall not go into here). 
Putting together eqs.~(\ref{deltatheta},\,\ref{deltaeta},\,\ref{deltaeta2},\,\ref{deltaT}) leads to the following result for the squared isocurvature fluctuations
\beq
\left\langle\left({\delta T\over T}\right)^{\!2}\right\rangle_{\!\!iso} \approx {9\,\gamf^2\over225\,\pi^2}{\Omega_b^2\over\Omega_m^2}{n^2H_i^2\over\rho_i^2}\cot^2(n\,\theta_i).
\label{isoresult}\eeq

We now compute the adiabatic temperature fluctuation. Recall from eq.~(\ref{DeltaR}) the formula for the adiabatic density fluctuations. To convert this to an adiabatic temperature fluctuation requires averaging over spherical harmonics, etc. The net result is roughly a factor of 1/20 difference, i.e.,
\beq
\left\langle\left({\delta T\over T}\right)^{\!2}\right\rangle_{\!\!adi} \approx {1\over20}{H_i^2\over 8\pi^2\mpl^2\epsilon_{sr}},
\label{adiresult}\eeq
(using $V_i\approx3H_i^2\mpl^2$). Since the isocurvature fluctuation is small, we can use this adiabatic fluctuation as an approximation for the total fluctuations.

The ratio of the isocurvature fluctuations to the total fluctuations is called $\alpha_{II}$. Putting together eqs.~(\ref{isoresult},\,\ref{adiresult}) we obtain
\beq
\alpha_{II}\approx {32\gamf^2\over5}{\Omega_b^2\over\Omega_m^2}{n^2\mpl^2\epsilon_{sr}\over\rho_i^2}\cot^2(n\,\theta_i).
\label{alpha1}\eeq
Planck data reveals that the baryon-to-matter ratio is $\Omega_b/\Omega_m\approx 0.16$. Lets take $\gamf\sim2$ as a representative value and $\cot(n\,\theta_i)\sim1$, this gives an isocurvature fraction
\beq
\alpha_{II}\sim 0.7{n^2\mpl^2\epsilon_{sr}\over\rho_i^2}.
\label{alpha2}\eeq
Note that this result holds for any simple inflationary potential. 

If we now specialize to the case of quadratic inflation, we have $\epsilon_{sr}\approx1/(2 N_e)$ and $\rho_i\approx2\sqrt{N_e}\,\mpl$. Then setting $N_e\approx 55$, we have our prediction for the isocurvature fraction
\beq
\alpha_{II}\sim 3\times 10^{-5}\,n^2.
\eeq
Recent Planck results have provided an upper bound on cold dark matter isocurvature fluctuations of \cite{Ade:2013uln}
\beq
\alpha_{II}<3.9\times 10^{-2},\,\,\,\,\,\,95\%\,\,\mbox{confidence},
\eeq
and we shall use this as a rough bound on baryon isocurvature fluctuations.
This suggests that only ridiculously large values of $n$ ($n>36$) are ruled out; but such values are unrealistic anyhow.
For the lowest value of $n$, namely $n=3$, we predict $\alpha_{II}\sim 3\times 10^{-4}$, i.e., two orders of magnitude below the current bound. 
On the other hand, in Section \ref{High} we explained that moderately high values of $n$ are especially interesting.
For instance, if we take $n= 8,\,10,\,12$, then our prediction is $\alpha_{II}\sim 3\times 10^{-3}$, i.e., only one order of magnitude below the current bound. This is quite exciting as it is potentially detectable in the next generation of data.

\section{Discussion and Conclusions}\label{Conclusions}

In this work we have proposed a way to directly unify early universe inflation and baryogenesis, with motivation from the Affleck-Dine mechanism. We developed in great detail the basic proposal summarized in our accompanying letter \cite{BaryogenesisLetter}. 

{\em Inflationary models:} As a concrete example, we studied the simplest inflation model; a quadratic (``chaotic") inflation potential. Other potential functions, such as hill-top models, cosine potentials (``natural inflation"), or other non-polynomial potentials which are concave down during inflation, are marginally preferred by recent CMB data \cite{Ade:2013uln}. Our methods are directly applicable to these cases and can be adopted straightforwardly. Our idea is simply to use a symmetric potential for inflation, under a $U(1)$ global symmetry, and then introduce a sufficiently weak breaking that does not spoil the flatness of the potential. This will generate a particle number during the latter stage of inflation. Other inflationary models, which  go far beyond this minimal inflationary setup, such as models dominated by higher derivative kinetic terms, appear to be disfavored for over predicting non-Gaussianity, etc. 

{\em Particle physics models:} We proposed two interesting particle physics models of this idea. The first model was to promote the $U(1)$ breaking to the level of a good symmetry until high dimension operators $n\geq8$. This allows $\phi$ to be a gauge singlet and then decay to quarks (and leptons) through dimension 7 operators, while still conserving baryon number. If the decay is controlled by the scale $\Lambda\sim 10^{15-16}$\,GeV, then we obtain good agreement with the observed baryon-to-photon ratio. Importantly, this value for $\Lambda$ is precisely in the regime of validity of the field theory; between the Hubble scale and the Planck scale.
The second model was to promote $\phi$ to a colored scalar, and allow decay to quarks readily through lower dimension operators. This requires a very small breaking parameter, which is technically natural, or a very small coupling to quarks, which deserves further exploration in contexts such as supersymmetry.

{\em Inflaton constraints:} An important parameter in the analysis is the inflaton mass $m$. As can be seen in eq.~(\ref{drcond}) the required decay rate to obtain the observed asymmetry scales as $\drreq\propto m^2$. So if $m$ is much smaller than the $m\sim 10^{13}$\,GeV used as a reference value in this work, then the required decay rates become much smaller. If decay occurs through high dimension operators then this is easily achieved, while if decay occurs through low dimension operators then this becomes more difficult. In simple models of inflation, the mass of the inflaton tends to be related, within an order of magnitude or so, to the Hubble parameter. So this intertwines parameters of high energy particle physics and the energy scale of inflation in an interesting way (complimentary to the difficulty at low energies \cite{Hertzberg:2011rc}). In particular, this means that decay through low dimension operators, as would be allowed by a colored inflaton, tends to favor high scale inflation. In turn this favors appreciable tensor modes, which are being actively searched for.

{\em Large scale dipole:} A distinguishing feature of these models, compared to other more common forms of baryogenesis, is that the dynamics respect the C and CP symmetry. Instead it is broken spontaneously by the initial state of the inflaton in the complex plane. For an initially inhomogeneous inflaton field, different regions of the universe will acquire different baryon-to-photon ratios in the late universe. This is an exciting property of the theory. This would allow for a large scale dipole in the baryon density in the universe, and could even be relevant to CMB anomalies \cite{Ade:2013nlj}. This is analogous to, and may be correlated with, a large scale dipole in the dark matter-to-photon density in the universe. Indeed the latter can occur if the dark matter is comprised of axions with a large Peccei-Quinn scale \cite{Tegmark:2005dy,Hertzberg:2008wr}. So if one were to observe a dipole in one or both of these densities, it would provide tremendous clues about the early universe and fundamental physics. Alternatively, these effects would be small if there were many e-foldings of inflation or if our pocket universe arose from bubble nucleation.

{\em Isocurvature -- prediction:} We also found that these models predict a baryonic isocurvature fluctuation at a level consistent with current observational bounds, and summarized in eqs.~(\ref{alpha1},\,\ref{alpha2}) for any symmetric potential $V_s$. For the quadratic inflation case we found an isocurvature fraction, one or two orders of magnitude below the current bounds; potentially detectable in the next generation of experiments.  For other choices of the symmetric potential $V_s$, the slow-roll parameter $\epsilon_{sr}$ will be different. This parameter is related to the tensor-to-scalar ratio by $r=16\,\epsilon_{sr}$. For simple models of inflation $\rho_i\sim\mbox{few}\,\mpl$, so all parameters are essentially fixed, leaving a relationship between the isocurvature fraction $\alpha_{II}$ and the tensor-to-scalar ratio $r$. An observational confirmation of this relationship would make our proposal quite compelling.

{\em Isocurvature -- comparison:}
Let us compare the aforementioned isocurvature fluctuation to the usual Affleck-Dine scenario where $\phi$ is not the inflaton. In that case, there is no obvious reason why the vev of the field during inflation $\rho_i=\sqrt{2}|\phi_i|$ must be larger than $\mpl$. Instead if $\rho_i$ is somewhat smaller, say of order the GUT scale, then the isocurvature fraction would be very large (at least for reasonably high scale inflation) and already ruled out \cite{Kasuya:2008xp}. By contrast, the current proposal of identifying $\phi$ with the inflaton naturally explains why $\rho_i$ is of the order of or slightly larger than $\mpl$, which is especially interesting.

{\em Small scales:} Another important subject is the possibility of inhomogeneity on very small scales. In this paper we focussed on the homogeneous mode of the inflaton (plus large scale inhomogeneities in Section \ref{Observational}). A potentially important consideration is the possibility of preheating after inflation \cite{Kofman:1997ga}. Under certain conditions, non-linear dynamics will lead to explosive production of high $k$ modes, and the breakup of $\phi$; possibly into objects such as Q-balls \cite{Coleman,Enqvist:1997si} and oscillons \cite{Copeland:1995fq,Amin:2010dc,Amin:2011hj}. If this is efficient it could throw the field up the potential in different ways in different patches of the universe. This means that the effective $\theta_i$ could be different in different patches of the universes on small scales. After decay, this would lead to patches of baryons and anti-baryons \cite{Dolgov:2008wu} which would presumably annihilate during thermalization, reducing the final value of $\eta$. On the other hand, this explosive process may raise the reheat temperature, raising the final value of $\eta$. So there are potentially competing effects. Such a process should only be important if the potential is strongly non-linear, which is not the case for quadratic inflation, or if there is significant couplings to other bosonic fields, such as $\sim\phi^*\phi\, H^\dagger H$. These considerations will be important for some inflationary models and is a topic of future work.

\vspace{0.2cm}
\begin{center}
{\bf Acknowledgments}
\end{center}
We would like to thank Mustafa Amin, Michael Dine, Alan Guth, Jesse Thaler, Tanmay Vachaspati, and Frank Wilczek for discussion, and we would like to acknowledge support by the Center for Theoretical Physics at MIT. This work is supported by the U.S. Department of Energy under cooperative research agreement Contract Number DE-FG02-05ER41360. JK is supported by an NSERC PDF fellowship.

\appendix

\section{Analytical Version of Affleck-Dine}\label{AppAffleckDine}

In Ref.~\cite{AffleckDine} Affleck-Dine studied a  simpler problem. They studied a scalar field in a (non-inflationary) background, with a known Hubble parameter, say $H=2/3t$ for matter era or $H=1/2t$ for radiation era, with some initial starting value for the field, say $\rho_i$. In order to solve this problem to order $\bp$ the authors used a perturbative technique, where they solved for the field at zeroth order in $\bp$ and then used this as a source term to solve for the field at first order in $\bp$, where the latter solution involves tracking the full complex field. This leads to an approximation for the asymmetry $\as$ in terms of some constants that they obtained numerically.

Here we mention that by using the techniques of Sections \ref{Asymmetry}, \ref{SmallGamma} we can solve this problem much more rapidly (only needing the zeroth order solution and solving only a single ordinary differential equation, rather than a coupled system) and we obtain the coefficients analytically.

At zeroth order the equation of motion for the field (after factorizing for $e^{i\theta_i}$) is
\beq
\ddot\rho_0+3H\dot\rho_0+m^2\rho_0=0,
\eeq
with $H$ specified by the background (here we will not rescale the field by $\mpl$ to make it dimensionless, since $\mpl$ is not relevant in this computation).
In the Affleck-Dine case one recognizes that the field is frozen at early times when $H\gg m$. So we impose initial conditions: $\dot\rho=0$ at early times. It is then easy to solve this differential equation. For the matter and radiation cases the solution is
\bea
&&\rho_0(t)={\sin(mt)\over mt}\rho_i,\,\,\,\,\,\,\,\,\,\,\,\,\,\,\,\,\,\,\,\,\,\,\,\,\,\,\,\,\,\,\,\,\,\,\,\,\,\,\,\,\mbox{Matter}\\
&&\rho_0(t)=2^{1/4}\,\Gamma(5/4){J_{1/4}(m\,t)\over (m\,t)^{1/4}}\rho_i,\,\,\,\,\,\,\,\mbox{Radiation}\,\,\,\,
\eea
where $\Gamma$ is the (complete) gamma function. 
This allows us to readily perform the integral that appears in $\Delta \Nphi$ (eq.~(\ref{deltaNsimp})) by taking the limits of integration $t_i=0$ and $t_f=\infty$. We write the scale factor in each case as
\bea
&&a=a_0\left(t\over t_0\right)^{2/3},\,\,\,\,\,\,\,\,\mbox{Matter}\\
&&a=a_0\left(t\over t_0\right)^{1/2},\,\,\,\,\,\,\,\mbox{Radiation}
\eea
in terms of some arbitrary reference constants $a_0,\,t_0$.
The integral that appears in $\Delta \Nphi$
\beq
I\equiv \int_0^\infty dt\,a(t)^3\rho_0(t)^n,
\eeq
can be easily performed in each of the eras. We find
\bea
&&I={a_0^3\rho_i^n\over t_0^2 m^3}b_n,\,\,\,\,\,\,\,\,\,\,\,\,\,\,\,\,\,\mbox{Matter}\\
&&I={a_0^3\rho_i^n\over t_0^{3/2}m^{5/2}}d_n,\,\,\,\,\,\,\,\mbox{Radiation}
\eea
where
\bea
&&b_n\equiv\int_0^\infty d\tau\,\tau^{2-n}\sin(\tau)^n,\\
&&d_n\equiv2^{n/4}\,\Gamma(5/4)^{n}\!\int_0^\infty d\tau\,\tau^{3/2-n/4}J_{1/4}(\tau)^n,\,\,\,
\eea
The first few values of these constants are
\bea
&&b_3={\pi\over4},\,\,\,\,b_4={\pi\over 4},\,\,\,\,b_5={5\pi\over32},\nonumber\\
&&b_6={\pi\over8},\,\,\,\,b_7={77\pi\over768},\,\,\,\,b_8={\pi\over12},\\
&&d_3={2\,\Gamma(5/4)^3\over 3^{1/4}\sqrt{\pi}\,\Gamma(3/4)}\approx 0.521,\nonumber\\
&&d_4={4\,\Gamma(5/4)^5\over\sqrt{\pi}\,\Gamma(3/4)^3}\approx 0.750.
\eea

In order to compute $\as$ we need to divide by the energy density $\varepsilon_0$ at late times. It is simple to show that at late times we have
\bea
&&\varepsilon_0={\rho_i^2\over 2t^2},\,\,\,\,\,\,\,\,\,\,\,\,\,\,\,\,\,\,\,\,\,\,\,\,\,\,\,\,\,\,\,\,\,\,\,\,\,\,\,\,\,\,\mbox{Matter}\\
&&\varepsilon_0={\sqrt{2}\,\Gamma(5/4)^2m^2\rho_i^2\over\pi (m\,t)^{3/2}},\,\,\,\,\,\,\,\mbox{Radiation}
\eea
Now recall  the definition of the asymmetry parameter $\as$ from eqs.~(\ref{deltaNsimp},\,\ref{density},\,\ref{ratio}). Putting the pieces together in the  Affleck-Dine regime, we obtain
\bea
&&\as=-\tilde{b}_n{\bp\rho_i^{n-2}\over m^2}\sin(n\,\theta_i),\,\,\,\,\,\,\,\,\,\,\mbox{Matter}\\
&&\as=-\tilde{d}_n{\bp\rho_i^{n-2}\over m^2}\sin(n\,\theta_i),\,\,\,\,\,\,\,\,\,\mbox{Radiation}
\eea
where
\bea
&&\tilde{b}_n\equiv{n\over 2^{n/2-2}}b_n,\\
&&\tilde{d}_n\equiv{\pi\,n\over 2^{(n-1)/2}\,\Gamma(5/4)^2}d_n.
\eea
This is in rough agreement with \cite{AffleckDine}, where the authors computed the constant prefactors for $n=4$ numerically using their alternate technique. Their numerics is found to be in small error $\sim 10\%-20\%$ from the correct values obtained analytically here.

\vspace{0.4cm}
{\em Email}: $^*$mphertz@mit.edu, $^\dagger$karoubyj@mit.edu

\end{document}